\documentclass[preprint,sort&compress]{elsarticle}
\usepackage{amsmath,amssymb}
\usepackage{lscape}
\usepackage{booktabs}

\begin{document}
\title{Search for Axion like particles using Laue-case conversion in a single crystal}
\author[phys]{T.~Yamaji}
\ead{yamaji@icepp.s.u-tokyo.ac.jp}
\author[sp8]{K.~Tamasaku}
\author[icepp]{T.~Namba}
\author[kek]{T.~Yamazaki}
\author[phys]{Y.~Seino}
\address[phys]{Department of Physics, Graduate School of Science, The University of Tokyo, 7-3-1 Hongo, Bunkyo, Tokyo 113-0033, Japan}
\address[sp8]{RIKEN SPring-8 Center, 1-1-1 Kouto, Sayo-cho, Sayo-gun, Hyogo 679-5148, Japan}
\address[icepp]{International Center for Elementary Particle Physics, The University of Tokyo, 7-3-1 Hongo, Bunkyo, Tokyo 113-0033, Japan}
\address[kek]{High energy Accelerator Research Organization, KEK, 203-1 Tokai-mura, Naka-gun, Ibaraki 319-1106, Japan}

\begin{abstract}
Axion Like Particles (ALPs) with a sub-keV range mass are searched by using the light-shining-through-a-wall technique. A novel system is developed in which injected X rays are converted and reconverted by the Laue-case conversion within a silicon single crystal with dual blades. The resonant ALPs mass of the conversion is scanned by varying the X-ray injection angle to the crystal. No significant signals are observed, and 90\% C. L. upper limits on the ALP-two photon coupling constant are obtained as follows,
\begin{eqnarray}
g_{a\gamma\gamma} &<& 4.2\times 10^{-3}~{\rm GeV^{-1}}~(m_a<10~{\rm eV}),\\
g_{a\gamma\gamma} &<& 5.0\times 10^{-3}~{\rm GeV^{-1}}~(46~{\rm eV}<m_a<1020~{\rm eV}).
\end{eqnarray} 
These are the most stringent laboratory-based constraints on ALPs heavier than 300~eV.
\end{abstract}

\maketitle

\section{Introduction}
Axion Like Particles (ALPs)~\cite{bib:ALPs1,bib:ALPs2,bib:ALPs3} predicted by theories beyond the Standard Model are also of astronomical interest because they can provide possible explanation for anomalous observations of the universe. ALPs are one of the viable candidates for the dark matter~\cite{bib:DM1,bib:DM2}. It has been also suggested that ALPs with a sub-keV range mass may be related to anomalous phenomena such as solar coronal heating and X rays from the dark side of the moon~\cite{bib:keV_motivation}.

ALPs have properties similar to an exotic particle, axion~\cite{bib:PQ1,bib:PQ2,bib:WW1,bib:WW2}, predicted by extended Peccei-Quinn Model of QCD never observed so far and able to solve the CP violation puzzle. ALPs and the original axion can interact with two photons via an anomaly diagram with fermions carrying exotic charges. The interaction mixes ALPs and photons under external electromagnetic fields, which is referred to as the Primakoff effect~\cite{bib:primakov1}. The interaction Lagrangian density can be represented as follows, 
\begin{eqnarray}
\mathcal{L}_{\rm int}=-\frac{g_{a\gamma\gamma}}{4}\mathcal{F}_{\mu\nu}\mathcal{\tilde{F}}^{\mu\nu}=g_{a\gamma\gamma}{\bf E\cdot B}a,
\end{eqnarray} 
where $g_{a\gamma\gamma}$ is the ALP-two photon coupling constant, $\mathcal{F}_{\mu\nu}\mathcal{\tilde{F}}^{\mu\nu}$ is the product of the electromagnetic field strength tensor and its dual, ${\bf E\cdot B}$ is the pseudo-scaler dot product of electromagnetic fields, and $a$ is the ALP field. The Primakoff effect takes place only when the electromagnetic fields of photons and external fields are parallel to each other. ALPs and the original axion can be searched by almost the same experimental techniques using the mixing effect. Although the original axion has the proportionality between the coupling constant and the axion mass~\cite{bib:KSVZ1,bib:KSVZ2,bib:DFSZ1,bib:DFSZ2}, the ALPs coupling constant is not bound to the ALP's mass, $m_a$. Searches for ALPs target the whole parameter region of $m_a$ and $g_{a\gamma\gamma}$.

Solar axion searches have imposed stringent upper limits on $g_{a\gamma\gamma}$ within a broad mass range~\cite{bib:SUMICO,bib:CAST,bib:COSME,bib:SOLAX,bib:DAMA,bib:CDMS}. However, these limits are inevitably model-dependent. The production rate of solar ALPs may be reduced by exotic models in which $m_a$ and $g_{a\gamma\gamma}$ depend on temperature and matter density within stellar systems~\cite{bib:reduction_model,bib:ALPs_reduction}. Laboratory-based experiments are scientifically important since they can complement more model-dependent celestial searches.

Model-independent laboratory-based searches have been performed by using the Primakoff effect under external magnetic fields~\cite{bib:BFRT1,bib:BFRT2,bib:BMV_LSW1,bib:BMV_LSW2,bib:GammaV,bib:LIPSS1,bib:LIPSS2,bib:ALPS0,bib:ALPS,bib:OSQAR0,bib:OSQAR,bib:LSWXE1,bib:LSWXE2,bib:NOMAD}. These experiments mainly utilize an experimental technique referred to as ``Light-Shining-through-a-Wall (LSW)"~\cite{bib:LSW1}. LSW experiments convert real photons from an laboratory source into ALPs under an external magnetic field. Unconverted photons are blocked by an opaque wall, and then ALPs are reconverted into photons in a second magnetic field. The sensitivity to $g_{a\gamma\gamma}$ is determined by the product of the magnetic field strength and length, which are typically $\mathcal{O}(1)$~T and $\mathcal{O}(1)$~m, respectively. The Primakoff effect under the magnetic fields efficiently generates and reconverts ALPs whose mass satisfies the following resonant condition, 
\begin{equation}
\left|m_a^2-m_\gamma^2\right|<\frac{4k_\gamma}{L_{\rm M}}
\label{eq:ma_m}
\end{equation}
where $m_\gamma$ is the plasma frequency of the media within the magnetic fields, $k_\gamma$ is the photon energy, and $L_{\rm M}$ is the magnetic field length. The ALPs mass satisfying the resonant condition (referred to as the resonant ALPs mass in this paper) has an upper limit determined by $m_\gamma$, $k_\gamma$, and $L_{\rm M}$, which are experimentally difficult to be changed continuously. The previous experiments search only ALPs with a mass of up to $\sim$40~eV~\cite{bib:NOMAD} since the plasma frequency and the photon energy cannot be increased arbitrarily. 

The Primakoff effect can take place also under electric fields such as high atomic electric fields within single crystals. The atomic electric fields are as high as $10^{11}$~V/m, which correspond to high magnetic fields of $10^{3}$~T. The Primakoff effect within single crystals has been explored by {\it B\"{u}chmuller and Hoogeveen}~\cite{bib:braggtheory}, {\it Liao}~\cite{bib:simpletheory}, and {\it Yamaji, et. al.}~\cite{bib:yamaji_laue}. The high electric fields can coherently convert injected X rays or ALPs with an energy of $\sim 10$~keV. The conversion probability can be represented as follows, 
\begin{eqnarray}
P_{a\rightarrow\gamma}&=&\left(\frac{1}{2}g_{a\gamma\gamma}E_TL_{\rm eff}{\rm cos}\theta_{\rm B}\right)^2,\label{eq:prob_laue}\\
L_{\rm eff}&=&2L_{\rm att}\left(1-{\rm exp}\left(-\frac{L}{2L_{\rm att}}\right)\right),
\end{eqnarray}
where $E_T$ is an effective electric field, $L_{\rm eff}$ is an effective conversion length, $\theta_{\rm B}$ is the Bragg angle, $L$ is the X-ray path length within the crystal, and $L_{\rm att}$ is the X-ray attenuation length. The effective conversion length is shorter than $L$ since the crystal absorbs X rays. The Bragg-case conversion, in which lattice planes are parallel to the crystal surfaces, has been utilized by solar axion searches~\cite{bib:COSME,bib:SOLAX,bib:DAMA,bib:CDMS}. It is recently shown in Ref.~\cite{bib:yamaji_laue} that $L_{\rm eff}$ for the Laue-case conversion, in which lattice planes are perpendicular to the crystal surfaces, is longer than that of the Bragg-case one. The effective conversion lengths of silicon crystals are $\sim 10^{-3}$~m, and the sensitivity to $g_{a\gamma\gamma}$ ($\propto E_T L_{\rm eff}$) is the same order as that of X-ray LSW experiments using external magnetic fields \cite{bib:LSWXE1, bib:LSWXE2}. 

The resonant ALPs mass of the Laue-case conversion is constrained by the condition as follows~\cite{bib:simpletheory,bib:yamaji_laue},
\newcommand{\ltsim}{\protect\raisebox{-0.5ex}{$\:\stackrel{\textstyle <}{\sim}\:$}} 
\begin{eqnarray}
\label{eq:resonant_mass}
\left|m_a^2-m_\gamma^2-2q_T\left(k_\gamma{\rm sin}\theta^\gamma_T-\frac{q_T}{2}\right)\right|\ltsim\frac{4k_\gamma}{L},
\end{eqnarray}
where $m_\gamma=\mathcal{O}(10)$~eV is the plasma frequency of the crystal, $q_T=\mathcal{O}(10)$~keV is the reciprocal lattice spacing, and $\theta^\gamma_T$ is an X-ray injection angle. Although the expression is almost the same as Eq.~(\ref{eq:ma_m}), an additional factor emerges on the left side. The factor within the parenthesis is equivalent to the deviation from the Bragg condition, $k_\gamma{\rm sin}\theta_{\rm B}=\frac{q_T}{2}$. This factor can be simplified as $k_\gamma\Delta\theta{\rm cos}\theta_{\rm B}$ in the case of $\Delta\theta\simeq 0$, where $\Delta\theta=\theta^\gamma_T-\theta_{\rm B}$ is the deviation of the X-ray injection angle from the Bragg angle (the detuning angle). The resonant ALPs mass can be scanned by rotating conversion crystals and varying $\Delta\theta$ since it is roughly proportional to $\sqrt{\Delta\theta}$. The scanning procedure can easily search massive ALPs (up to $10$~keV) in comparison with the previous LSW experiments. 

This paper reports the first LSW experiment using the Laue-case conversion within a single crystal. The experiment utilizes BL19LXU beam line of SPring-8. Injected X rays are converted and reconverted by a novel system composed of thin silicon blades and an opaque wall. The resonant ALPs mass is continuously scanned by rotating the conversion system.

\section{Experimental setup and measurement}
The experiment has been performed during a 96-hour beam time of BL19LXU in Oct. 2017. SPring-8 is a third-generation synchrotron radiation facility producing nearly continuous X-ray beams. BL19LXU beam line is the strongest beam line among the facility~\cite{bib:bl19}. The undulator of BL19LXU radiates horizontally-polarized X rays.

\begin{figure}[!t]
  \centering
  \includegraphics[angle=0,width=1.0\textwidth]{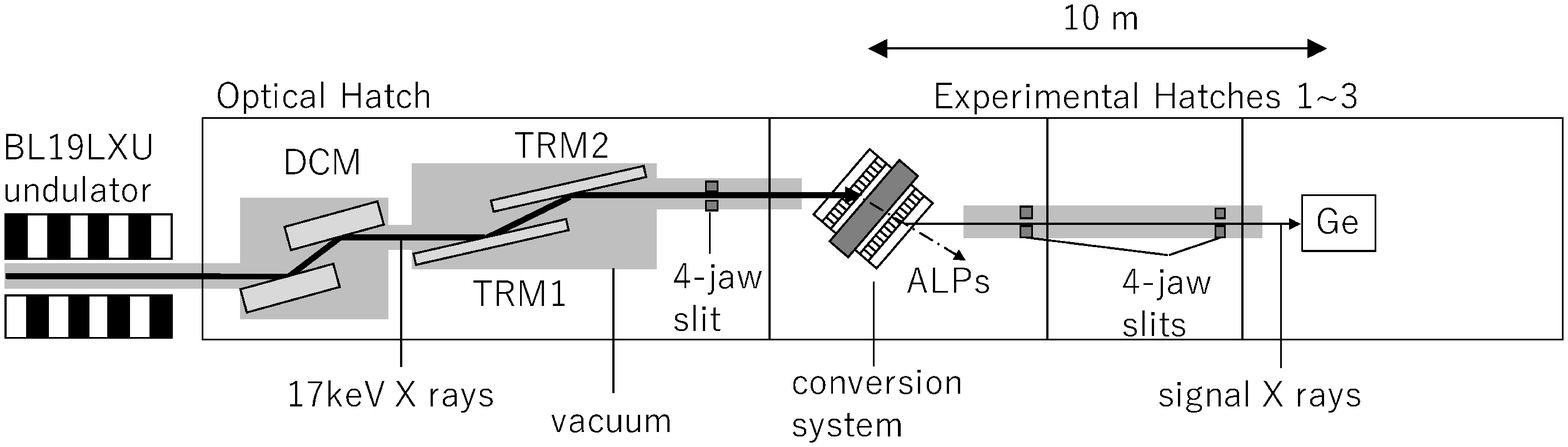}
  \caption{Schematics of the whole experimental setup. The shaded regions are evacuated to avoid the X-ray absorption due to atmospheric molecules.}  
  \label{fig:optics}
\end{figure}

The whole experimental setup is shown in Fig.~\ref{fig:optics}. The photon energy of fundamental radiation is tuned to be 17~keV in order for the X-ray attenuation length to accord with the thickness of silicon blades shown later. The X-ray beams are monochromatized to a bandwidth of 2.1~eV (FWHM) by a double-crystal monochromator (DCM) located at the optical hatch. Higher-harmonics are rejected by a total reflection mirror system (TRMs) to enhance the signal-to-noise ratio of the experiment. The X-ray beams are collimated by a tantalum four-jaw slits with an opening window of $2\times 2~{\rm mm^2}$ to block stray X rays. The flux, the angular divergence, and the widths of injected X rays are $2.5\times 10^{13}$~photons/s, $6.1~{\rm \mu rad}$ (FWHM), and $0.5 ({\rm V})\times 0.6 ({\rm H})~{\rm mm^2}$ (FWHM), respectively.

\begin{figure}[!t]
  \centering
  \includegraphics[angle=0,width=1.0\textwidth]{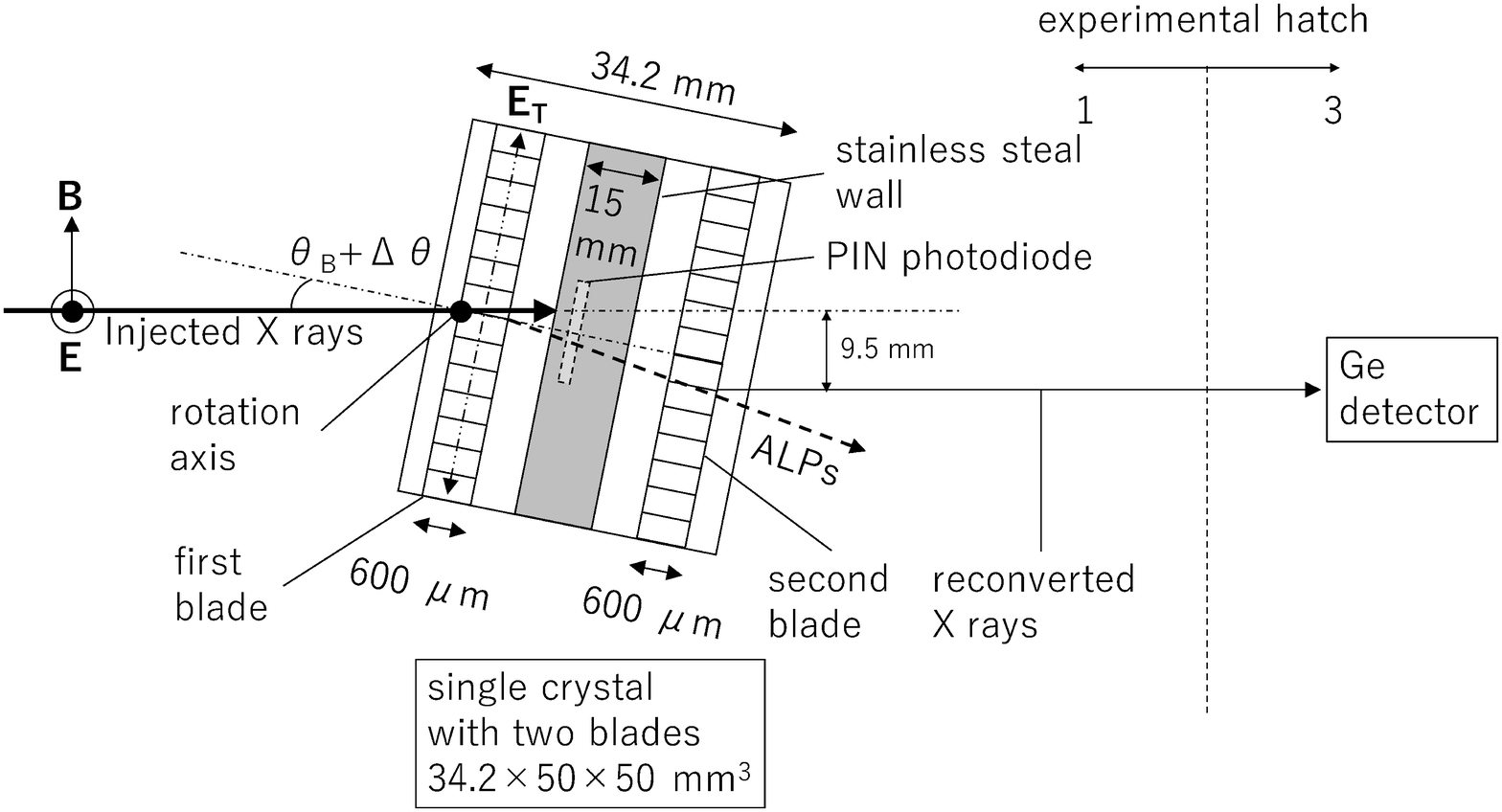}
  \caption{Schematics of the conversion system.}  
  \label{fig:system}
\end{figure}

The conversion system is installed into the experimental hatch 1. The system is composed of a silicon single crystal with dual blades and an opaque wall as shown in Fig.~\ref{fig:system}. The system is covered by HDPE sheets to reduce the fluctuation of the temperature around the system. The silicon single crystal has two $600~{\rm \mu m}$-thick blades manufactured on it. The crystal has Si(220) lattice planes perpendicular to the blade surfaces (Laue-case). The effective electric field of Si(220) is calculated to be $E_T=4.1\times 10^{10}$~V/m (the Dirac-Fock method~\cite{bib:DF, bib:DIRAC}). The reciprocal lattice spacing and the Bragg angle are $q_T=6.46$~keV and $\theta_{\rm B}=10.95$~degrees, respectively, in this case. The X-ray attenuation length is $L_{\rm att}=650~{\rm \mu m}$ for 17~keV X rays, and the effective conversion length is calculated to be $L_{\rm eff}=488~{\rm \mu m}$. The value of $E_TL_{\rm eff}{\rm cos}\theta_{\rm B}$ corresponds to $0.23~{\rm GeV}$ in the natural unit. The opaque wall is made of a 15~mm-thick stainless steel plate, which is equipped with a PIN photodiode (HAMAMATSU S3590-09) to monitor the intensity of injected X rays. The X-ray injection angle is aligned by using a precision goniometer, KOHZU RA20-21. The goniometer is driven discretely by a stepping motor with a stepping angle of $\Delta\theta_{\rm step}=0.17~{\rm \mu rad}$. 

The first blade converts injected X rays with $\sigma$-polarization into ALPs vertically downwards, and these ALPs are reconverted into X rays (signal X rays) by the second blade. The signal X rays are parallel to the injected ones and have the same X-ray photon energy of 17~keV. The bandwidth of the resonant ALPs mass, Eq.~(\ref{eq:resonant_mass}), is equivalent to an acceptable divergence of the X-ray/ALPs injection angles as follows,
\begin{equation}
\Delta\theta_{\rm CV}=\frac{2d_{220}}{\pi H}=\frac{2\times 1.92~{\AA}}{\pi \times 600~{\rm \mu m}}=204~{\rm nrad},
\end{equation}
where $d_{220}$ is the spacing of (220) lattice planes and $H$ is the crystal thickness. The parallelism between lattice planes within these blades is required to be less than $\Delta\theta_{\rm CV}$ in order to effectively reconvert ALPs into X rays. The parallelism is guaranteed with the precision of $\sim 1$~nrad since these blades are made of the same single crystal. 

A germanium semiconductor detector, CANBERRA GL0210 with a crystal geometry of $\phi$16~mm$\times$t10~mm \cite{bib:germanium}, is installed into the experimental hatch 3 to suppress stray X rays scattered by the conversion system. The evacuated pipe between the conversion system and the detector is equipped with two four-jaw slits similar to the one within OH in order to suppress stray X rays further. The pulse height of the detected X rays is enhanced by an amplifier (ORTEC 572) and recorded by using a peak hold ADC (HOSIN C011) with an internal trigger. The energy resolution and the detection efficiency of the detector for signal X rays are measured to be $(111\pm 2)$~eV ($1\sigma$) and $\epsilon_d=(82\pm1)$\%, respectively, by using radioisotopes ($^{55}$Fe, $^{57}$Co, and $^{241}$Am). The gain stability of the detector during the beam time is also measured to be $\pm10$~eV by using 13.95~keV and 17.75~keV X rays from $^{241}$Am. The energy window for signal X rays is determined to be ($17.00 \pm 0.23)~{\rm keV}$ by taking into account the gain stability and the resolution ($2\sigma$). The detector is covered almost all around by 5~cm-thick lead blocks to reduce environmental X rays. The detection rate of environmental X rays within the energy window is measured to be $(0.51\pm 0.10)\times 10^{-3}$~photons/s.

The detector position is aligned by using X rays reflected twice by the conversion system. Injected X rays can be reflected by Laue-case X-ray diffraction within the first blade under the Bragg condition. The reflected X rays are subsequently reflected by the second blade when the wall is opened. The X-ray beams reflected twice (RR beams) are almost coaxial to the signal X rays. The height difference between the RR beams and the signal X rays is an increasing function of $\Delta\theta$ ($m_a$), and the difference is $6~{\rm \mu m}$ for ALPs with a mass of 1~keV. The X-ray beam profile is also estimated by measuring the diffraction efficiency of the RR beams as a function of $\Delta\theta$, which is referred to as the rocking curve. The diffraction efficiency of the RR beams depends on the deviation from the Bragg condition in the same way as the Laue-case conversion. Since the Bragg condition couples the X-ray injection angle and the X-ray photon energy, the energy-angular profile of injected X rays is equivalent to the one with no bandwidth and an effective angular profile, $\frac{dF}{d\theta_X}$, for X-ray diffraction and the Laue-case conversion, where $F$ is the normalized profile of injected X rays and $\theta_X$ is the effective angular deviation from the center of injected X rays. The rocking curve has a profile expanded by X-ray diffraction from $\frac{dF}{d\theta_X}$. The expected $\Delta\theta$ distribution of signal X rays has the same profile as $\frac{dF}{d\theta_X}$ with a mean value corresponding to the ALP's mass. The signal-to-noise ratio also depends on $\frac{dF}{d\theta_X}$ since environmental X rays within the energy window and a $\Delta\theta$ window shown later are indistinguishable from signal X rays. 

\begin{figure}[!t]
  \centering
  \includegraphics[angle=0,width=0.6\textwidth]{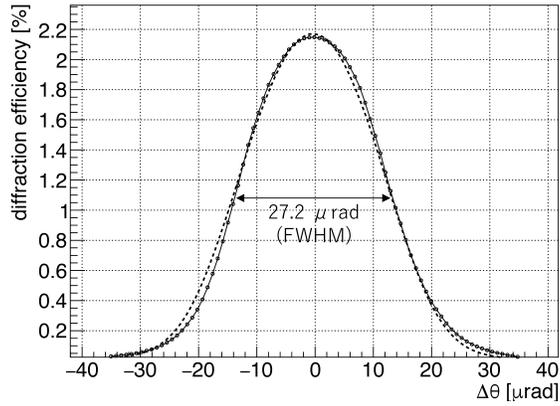}
  \caption{The measured rocking curve with the largest FWHM among the ones measured during the beam time (the solid line). The circles show data points. The calculated rocking curve is also shown by the dashed line.} 
  \label{fig:rocking}
\end{figure}

The total throughput of the reflection (two times) is measured by a S3590-09 PIN photodiode installed in front of the germanium detector in the experimental hatch 3. The measured rocking curve with the largest FWHM is shown in Fig.~\ref{fig:rocking}. The peak value and the width of the rocking curve are ($2.15\pm 0.01$)\% and $\Delta\theta_{\rm BL}=27.2~{\rm \mu rad}$ (FWHM, with the fluctuation of $<0.7~{\rm \mu rad}$), respectively. The measured width is larger than the angular divergence of injected X rays ($6.1~{\rm \mu rad}$) due to its bandwidth. These measured values are consistent with ones calculated by assuming a simple Gaussian energy-angular profile within the relative precision of $2$\%, which guarantees that X-ray diffraction within the conversion system is well understood and that the apparatus is correctly setup. The deviation between them can be attributed to the Gaussian approximation of the beam profile. The effective angular profile is approximated by the measured rocking curve in this paper. The approximation overestimates the effective angular divergence since the measured rocking curve is a convolution between the angular divergence and the intrinsic rocking curve. 

The goniometer angle is varied at $\Delta\theta_{\rm step}$ intervals from the Bragg angle during the measurement of signal X rays. The scanning range is set to be from $\Delta\theta=0~{\rm mrad}$ to $\Delta\theta=4.9~{\rm mrad}$, which corresponds to the resonant ALP's mass of $m_a<1$~keV. The data takings are performed four times with a scanning speed of 5.82, 5.66, 9.70, and 4.62~${\rm \mu rad/min}$. Two beam dumps take place during the third (at $\Delta\theta$=$3.99~{\rm mrad}$) and the fourth ($3.17~{\rm mrad}$) measurement. The overall data acquisition time is 47.2 hours. The fluctuation of X-ray intensity during the data takings is measured to be $<1.2$\% by the PIN photodiode within the wall. The goniometer angle drifts slowly due to the fluctuation of the room temperature. The drift angle, $\Delta\theta_{\rm D}$, is estimated by rocking curves measured before and after the data takings. The drift angles of the four data takings are $-15.7$, $-6.11$, $-10.8$, and $-12.9$~${\rm \mu rad}$, respectively. The X-ray injection angle also temporally shifts during the drive of the stepping motor by $\Delta\theta_{\rm DT}=+3.49~{\rm \mu rad}$ due to the distortion of a rotation coupler between the goniometer axis and the silicon crystal holder. Although these drifts are much smaller than the scanning range and do not affect the experimental sensitivity strongly, we take into account them in the signal integration procedure described later.

\begin{figure}[!t]
  \centering
  \includegraphics[angle=0,width=1.0\textwidth]{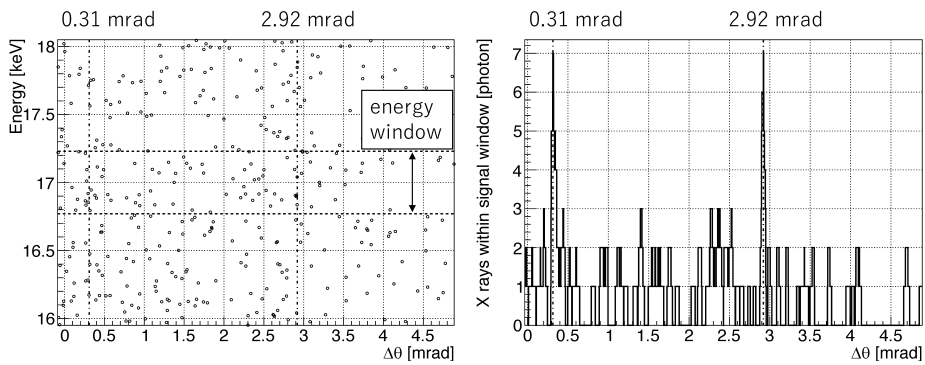}
  \caption{The $\Delta\theta$ distribution of signal X rays. Left: the energy-$\Delta\theta$ distribution of detected X rays in the vicinity of the energy window (scatter plot). The positions of peaks with seven photons are shown by the dash-dotted lines. Right: the $\Delta\theta$-integrated distribution of X rays within the energy window.}  
  \label{fig:compare}
\end{figure}

Figure~\ref{fig:compare} (left) shows the energy-$\Delta\theta$ distribution of detected X rays in the vicinity of the energy window. The sparse distribution outside the energy window shows the contribution of environmental X rays independent of $\Delta\theta$. The number and the rate of detected X rays within the energy window are 94~photons and $(0.554\pm0.057)\times 10^{-3}$~photons/s, respectively. The latter is consistent with the rate of the environmental X rays. The signal yield can be enhanced by integrating the detected X rays within a $\Delta\theta$ window due to the effective angular profile. The $\Delta\theta$ window for $\Delta\theta=\Delta\theta_i$ is defined as [$\Delta\theta_i-\Delta\theta_{\rm BL}/2-|\Delta\theta_{\rm DT}|$,~$\Delta\theta_i+\Delta\theta_{\rm BL}/2+|\Delta\theta_{\rm D}|$] where all contributions discussed above are taken into account. The definition of the $\Delta\theta$ window is conservative since the effective angular divergence is overestimated as shown above. Figure~\ref{fig:compare} (right) shows the integrated $\Delta\theta$ distribution of detected X rays. The spectrum has two fine peaks with seven photons at 0.31/2.92~mrad, where a mean number of environmental X rays is expected to be $(1.0\pm 0.1)$~photons from the measured rate of environmental X rays. The peaks can be explained by the accidental accumulation of environmental X rays. The probability that environmental X rays make a peak higher than six photons is calculated to be $4.4$\% ($\sim$1.7$\sigma$) by assuming the Poisson distribution. The measured peaks are not significant enough to be identified as the signals of ALPs.

The upper limits on $g_{a\gamma\gamma}$ can be obtained as a function of $m_a$ from the effective number of injected X rays shown below. The number of X rays satisfying the resonant condition at $\Delta\theta=\Delta\theta_i$ can be approximated to be $N_\gamma^i\Delta\theta_{\rm CV}\frac{dF}{d\theta_X}\left(\theta_X=\Delta\theta_j-\Delta\theta_i\right)$, where $N_\gamma^i$ is the total number of injected X rays at $\Delta\theta=\Delta\theta_i$ and $\Delta\theta_j$ is the detuning angle corresponding to an ALPs mass. The number integrated within the $\Delta\theta$ window, $N_{\rm eff}^i$, can be interpreted as an effective number of injected X rays which can be converted into ALPs with a certain mass. $N_{\rm eff}^i$ can be represented as follows by assuming that $N_\gamma^i$ is constant within the $\Delta\theta$ window, 
\begin{eqnarray}
N_{\rm eff}^i=N_\gamma^i\frac{\Delta\theta_{\rm CV}}{\Delta\theta_{\rm step}}\sum_{|\theta_X|<\frac{\Delta\theta_{\rm BL}}{2}}\left[\frac{dF}{d\theta_X}\Delta\theta_{\rm step}\right]
\end{eqnarray}
where the summation is performed with the same stepping as $\Delta\theta$ ($\theta_X=n\Delta\theta_{\rm step}$). The values of $N_{\rm eff}^i$ are about $1.6\times 10^{14}$ ($\Delta\theta<3.17$~mrad), $1.3\times 10^{14}$ ($3.17~{\rm mrad}<\Delta\theta<3.99$~mrad), and $8.1\times 10^{13}$ photons ($\Delta\theta>3.99$~mrad), respectively. 

The upper limit can be simplified as follows by assuming that the conversion-reconversion probability has a constant value shown in Eq.~(\ref{eq:prob_laue}) within the required angular precision,  
\begin{eqnarray}
g_{a\gamma\gamma}&<&\left(\frac{1}{2}E_TL_{\rm eff}{\rm cos}\theta_{\rm B}\right)^{-1}\left(\frac{N^s_{i,CL}}{N_{\rm eff}^i\epsilon_d}\right)^{\frac{1}{4}}\\
{\rm for}~m_a&=&\sqrt{m_\gamma^2+2q_T\left\{k_\gamma{\rm sin}(\theta_{\rm B}+\Delta\theta_i)-\frac{q_T}{2}\right\}}.
\label{eq:sense}
\end{eqnarray}
$N^s_{i,CL}$ is an upper limit on the number of X rays within the signal window, which is calculated from the integrated $\Delta\theta$ distribution of detected X rays by assuming the Poisson distribution. The upper limit can be obtained as a function of $m_a\propto\sqrt{\Delta\theta}$.  

Systematic uncertainties related to $g_{a\gamma\gamma}$ and $m_a$ are summarized in Tab.~\ref{tab:uncertainties}. The largest uncertainties come from the estimation of $L_{\rm eff}$ and the effect of X-ray heat load. Minor X-ray diffraction on lattice planes other than Si(220) takes place during the scanning at $\Delta\theta=3.67~{\rm mrad}$. The X-ray diffraction reduces the X-ray attenuation length at $\Delta\theta>\sim 3.67~{\rm mrad}$. The sensitivity to $g_{a\gamma\gamma}$ is reduced by $3.7\%$ at $3.40~{\rm mrad}< \Delta\theta <3.84~{\rm mrad}$ and $1.4\%$ at $3.84~{\rm mrad}<\Delta\theta<4.9~{\rm mrad}$, where these values are estimated by the monitored X-ray intensity. The X-ray heat load on the first blade ($\sim 41.5~{\rm mW}$) reduces the Bragg angle at the conversion by expanding the crystal lattice. The effect on the Bragg angle is evaluated by a finite-element-method simulation code, ANSYS~\cite{bib:ANSYS}. The simulation ignores heat conduction via the atmosphere and assumes the X-ray beam profile to be rectangular with the same widths as the FWHM of injected X rays. These approximations overestimate the effect of the X-ray heat load. The shift of the Bragg angle is estimated to be $<97.3$~nrad, which reduces the reconversion efficiency and the sensitivity to $g_{a\gamma\gamma}$ by $<17.4$\% and $<4.4\%$, respectively.

The uncertainties shown in Tab.~\ref{tab:uncertainties} are linearly accumulated, and the sensitivities to $g_{a\gamma\gamma}$ and $m_a$ are deteriorated by $1\sigma$ of the overall uncertainty, as much as 9.5\% ($g_{a\gamma\gamma}$, max) and $0.33\%$ ($m_a$). The upper limits are numerically calculated as shown in Fig.~\ref{fig:limits}, and their values can be represented by the maximum upper limits shown below, 
\begin{eqnarray}
g_{a\gamma\gamma} &<& 4.2\times 10^{-3}~{\rm GeV^{-1}}~({\rm for}~m_a<10~{\rm eV}),\\
g_{a\gamma\gamma} &<& 5.0\times 10^{-3}~{\rm GeV^{-1}}~({\rm for}~46~{\rm eV}<m_a<1020~{\rm eV}).
\end{eqnarray}
The sensitivity around the plasma frequency of silicon, $m_\gamma=31$~eV, is reduced due to the effect of the Laue-case diffraction \cite{bib:yamaji_laue}. X rays propagate through the crystal as two standing waves, the Bloch waves $\alpha/\beta$, under the Bragg condition. The plasma frequencies of the Bloch waves are shifted from $m_\gamma$, and their contribution to the conversion of ALPs with a mass of $\sim m_\gamma$ interferes destructively. Figure~\ref{fig:limits} also shows upper limits imposed by previous LSW experiments. The obtained upper limits provide the most stringent laboratory-based constraint on ALPs heavier than $m_a=300$~eV. The sensitivity to ALPs heavier than resonant ALPs mass has large uncertainties since the sensitivity fluctuates rapidly as a function of experimental parameters. The obtained limits are also the first rigid constraints in the range of $40~{\rm eV}<m_a<1~{\rm keV}$. 

\section{Conclusion}
We have performed a novel LSW search using the Laue-case conversion in a silicon crystal. The resonant ALPs mass is continuously scanned by rotating the conversion system and varying the detuning angle. The obtained upper limits are the most stringent conditions on ALPs with a mass of $300~{\rm eV}<m_a<1~{\rm keV}$ as a laboratory-based search. The experimental results can be also considered as the first rigid constraints on ALPs heavier than $\sim$40~eV, where previous experiments using external magnetic fields cannot resonantly convert ALPs.

\section{Acknowledgements}
The research is funded by the Japan Society for Promotion of Science (Grant number 15J00509), and the experiment at BL19LXU is approved by RIKEN with the proposal number of 20170021. We would like to thank Shoji Asai for useful discussion and suggestions.

\bibliographystyle{phreport}
\bibliography{bibliography}

\clearpage

\begin{figure}
  \centering
  \includegraphics[width=0.6\textwidth, angle=0]{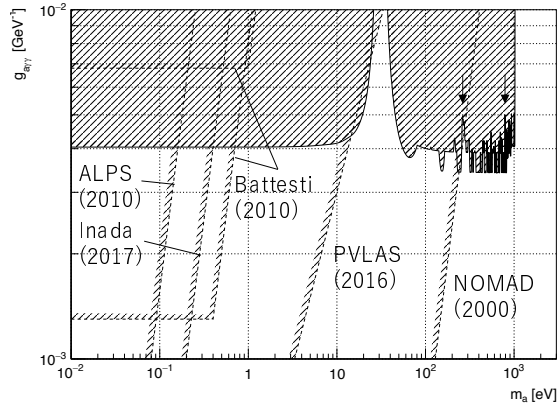}
    \caption{The comparison between the obtained limit and previous laboratory-based experiments. The exclusion region obtained by this experiment is shown as the hatched area. The upper limits obtained by previous experiments are shown by the dashed lines. The ALPs masses corresponding to the peaks with seven photons are shown by the arrows. The figure shows only laboratory-based experiments sensitive to ALPs with $m_a>10^{-2}$~eV. ALPS experiment~\cite{bib:ALPS} is an optical LSW experiment. Battesti (2010)~\cite{bib:LSWXE1} and Inada (2017)~\cite{bib:LSWXE2} show X-ray LSW experiments performed at ESRF and SPring-8. PVLAS experiment~\cite{bib:PVLAS} is a VMB experiment. NOMAD experiment~\cite{bib:NOMAD} uses high energy photons from a neutrino beam line. }
   \label{fig:limits}
\end{figure}

\begin{table}[!htbp]
  \begin{center}
    \small
    \begin{tabular}{lccc}
\toprule
uncertainty & affected factor & uncertainties on $g_{a\gamma\gamma}$ & $m_a$ \\
\midrule
blade thickness & $L_{\rm eff}$/$\Delta\theta_{CV}$ & $\pm 0.48\%$ & \\
the beam intensity & $N_\gamma^i$ & $\pm 0.33\%$ & \\
stray X rays from TRMs & $N_\gamma^i$ & +0.18\% & \\
drift of $\Delta\theta$& $N_\gamma^i/\Delta\theta_i$ & $\pm 0.08\%$ & $\pm 0.15\%$ \\
absolute accuracy of $\Delta\theta$ & $\Delta\theta_i$ & & $\pm 0.18\%$\\
detector efficiency & $\epsilon_d$ & $\pm 0.37\%$  & \\
accidental X-ray diffraction & $L_{\rm eff}$ & $0/3.7/1.4\%$ & \\
X-ray heat load & $P^2$ & $4.4\%$  &  \\
\midrule
Overall (conservative) & & $9.5$\% (max) & 0.33\%\\
\bottomrule
    \end{tabular}
  \end{center}
  \caption{Summary of systematic uncertainties on $g_{a\gamma\gamma}$ and $m_a$.}
  \label{tab:uncertainties}
\end{table}

\end{document}